# Self-induced polarization tracking, tunneling effect and modal attraction in optical fiber

M. Guasoni[1], P. Morin[1], P.-Y. Bony[1], S. Wabnitz[2] and J. Fatome[1,*]

[1] Laboratoire Interdisciplinaire Carnot de Bourgogne, UMR 6303 CNRS - Université Bourgogne Franche-Comté
, 9 avenue Alain Savary, 21078 Dijon, France.
[2] Department of Information Engineering, Università di Brescia, Via Branze 38, 25123 Brescia, Italy

*Corresponding author: jfatome@u-bourgogne.fr

In this paper, we report the observation and exploitation of the capability of light to self-organize its state-of-polarization, upon propagation in optical fibers, by means of a device called Omnipolarizer. The principle of operation of this system consists in a counter-propagating four-wave mixing interaction between an incident signal and its backward replica generated at the fiber output thanks to a reflective fiber loop. We have exploited this self-induced polarization tracking phenomenon for all-optical data processing and successfully demonstrated the spontaneous repolarization of a 40-Gbit/s On-Off keying optical signal without noticeable impairments. Moreover, the strong local coupling between the two counter-propagating waves has also revealed a fascinating aspect of the Omnipolarizer called polarization-based tunneling effect. This intrinsic property enables us to instantaneously let "jump" a polarization information onto the reflected signal, long before the expected time-of-flight induced by the round-trip along the fiber span. Finally, we discuss how the concept of self-organization could be generalized to multimode fibers, which paves the way to new important applications in the framework of spatial-mode-multiplexing.

## I. INTRODUCTION

Among the three independent features that characterize a light beam propagating in a monomode optical fiber, namely, the frequency, energy and state-of-polarization (SOP), the SOP remains the most elusive variable which is still difficult to predict and control. Outstanding technological developments in the manufacturing process of standard monomode optical fibers have been realized in the past decade. Especially, the implementation of a fast spinning process during the drawing stage now enables fiber providers to deliver standard telecom fibers with spectacular weak levels of polarization-mode dispersion [1-4]. Nevertheless, the residual random birefringence associated with mechanical stress, bending, squeezing, vibrations or temperature variations make the SOP of a light beam totally unpredictable after a few dozens of meters of propagation [5-10]. However, from a general point of view, despite the recent tremendous technological developments in waveguide and fiber-based systems to mitigate polarization impairments, the basic principle of operation of the associated SOP control methods often rests upon a combative strategy rather than on a preventive strategy. For instance, in high-capacity coherent transmissions, polarization impairments such as polarization randomness, polarization-mode dispersion [11-15], polarization depending loss [16] or cross-polarization interactions [17-18] are efficiently managed at the receiver by means of digital signal processing [19-21]. As far as highly polarization dependent systems such as on-chip integrated optical circuits or fiber-based nonlinear processing devices, special designs and more or less complex polarization-diverse schemes (polarization diversity, bi-directional loop or polarization splitting/recombination) may ensure the mitigation of polarization-dependent performances [22-25].

In order to master or control the SOP of light in fiber-based systems, the most effective strategy consists in implementing an opto-electronic polarization tracker [26-30]. Such devices are generally based on linear polarization transformations followed by partial diagnostic combined with an active feedback loop control driven by complex algorithms. Thanks to this well-established technique, record polarization tracking speeds have been achieved, reaching several of Mrad/s for commercially available units [31-34].

On the other hand, in order to benefit from all-optical alternatives for the development of future transparent networks, nonlinear effects have emerged in the last decade as a possible way to all-optically master the polarization of light propagating in optical fibers. To this aim, several techniques have emerged in the literature in order to develop a nonlinear polarizer capable to repolarize an incident signal with 100% of efficiency, whilst preserving the quality of the temporal intensity profile. This phenomenon of polarization attraction in optical fibers or polarization pulling effect, has been the subject of numerous studies in the literature involving the Raman effect [35-43], the stimulated Brillouin backscattering [44-49], the parametric amplification [50-52] as well as a counter-propagating four-wave mixing process, also called nonlinear cross-polarization interaction [53-68]. For this last particular case, it has been shown that an arbitrarily polarized incident signal can be attracted toward a specific SOP, which is fixed by the polarization of the counter-propagating pump wave injected at the opposite end of the fiber [55]. This phenomenon of polarization attraction has been the subject of numerous studies, reporting the efficient repolarization of telecom signals at 10 and 40 Gbit/s, in combination with several types of optical functionalities by using the same span of fiber, e.g., intensity profile regeneration for On/Off keying (OOK) formats [61], noise cleaning [62], data packet processing [63], Raman amplification [64], spatial mode attraction [65]. Nevertheless, the common feature of all of these previous works is that the injection of an external reference pump wave is a prerequisite for the existence of the polarization attraction process.

At the opposite of this general rule, our recent experimental observations have demonstrated that a spontaneous polarization attraction process can also occur in the absence of any SOP reference beam in a device called Omnipolarizer [66]. In this novel solution, the signal beam interacts with its own counter-propagating replica generated at the fiber end thanks to a single reflecting component, e.g., Fiber Bragg-Mirror (FBG), coating or amplified reflective fiber loop setup [66-68]. In this particular case, the signal itself evolves in time towards a stationary state imposed by this self-induced nonlinear polarization attraction process. The aim of this paper is thus to provide a general overview of this phenomenon, as well as to highlight some new results and discuss future developments.

The paper is organized as follows: in the first two sections, we introduce the principle of operation and theoretical description of the Omnipolarizer. Then, we present in more details the experimental demonstration of self-induced polarization tracking of a 40-Gbit/s OOK signal reported in ref. [66], enabling an error-free detection beyond a polarizer. In the second part of the manuscript we highlight a novel intriguing behavior of the Omnipolarizer, based on the strong local coupling between the two counter-propagating waves, which allows us to instantaneously induce a "jump" of polarization information onto the reflected signal, long before the expected time-of-flight into the fiber span, which may be described as a polarization-based tunneling effect. Finally, in the last sections we also discuss new perspectives for the generalization of the idea of self-organization to multimode fibers, and trace out our conclusions.

## II. Principle Of Operation

The principle of operation of the Omnipolarizer is depicted in Fig. 1. The device basically consists in a few-km long standard optical fiber encapsulated in between an optical circulator at the input and a reflective element at the opposite end. In this configuration and in order to observe a self-induced repolarization process, an arbitrary polarized incident signal has to nonlinearly interact with its backward replica. For an efficient cross-polarization effect along the entire fiber length, the counter-propagating beams should propagate at least some few nonlinear lengths, where the nonlinear length reads as $L_{nl}=1/\gamma P$, with $\gamma$ the nonlinear Kerr coefficient of the fiber and $P$ the input power [69]. Therefore, in a typical configuration involving a 5-km long standard optical fiber with a Kerr coefficient $\gamma \approx 2$ W$^{-1}$.km$^{-1}$, a relatively high level of power close to 500 mW is necessary.

Secondly, the Omnipolarizer can be characterized by its reflection coefficient $\rho$, which is defined as the ratio of power between backward and forward signals. Depending on the value of $\rho$, which has to be above 0.8 for an efficient interaction and can be even larger than 1 for specific purposes, two main operating regimes have been identified.

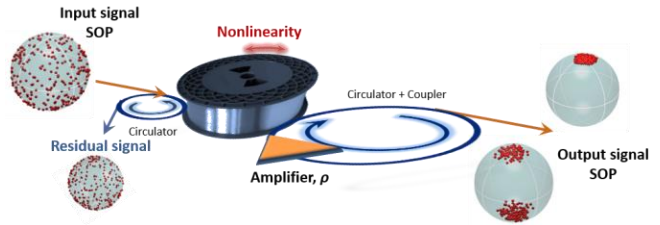

Fig. 1. Principle of operation of the Omnipolarizer.

The first operating regime is the *bistable* regime, and is reached for a reflection coefficient below unity, typically $0.8 \leq \rho \leq 1$. Basically, it corresponds to a simple reflection at the fiber end, and thus can be also achieved by means of a mirror, a FBG, a special coating or an amplified reflective fiber loop with a moderate amplifier gain. In this regime, for any arbitrarily polarized input signal, two opposite poles of attraction for the output SOP can be identified on the Poincaré sphere (see Fig. 1). In practice, the sign of the input signal ellipticity defines which of the two SOPs is obtained in the output. Consequently, for an initially depolarized signal, i.e., quickly scrambled in time all over the Poincaré sphere, the output signal SOP distribution will be highly localized around both poles of the sphere at the fiber output (Fig. 1) [68].

In the second operating regime, the backward signal is amplified in such a way to achieve a reflective coefficient in the range of $1.2 \leq \rho \leq 2$. In this case, any arbitrarily polarized input signal is attracted towards a single output SOP, whose position over the Poincaré sphere can be controlled by means of the polarization rotation imposed by the feedback loop (practically by means of a classical polarization controller implemented into the reflective loop) [66-67]. Given that any input SOP vector over the Poincaré sphere is aligned to a unique SOP vector at the device output, we define this functionality as the *alignment* regime.

Finally, for larger reflection coefficients, i.e., $\rho \gg 1$, a chaotic dynamics can be reached, leading to an all-optical scrambling of the output polarization [70], which is briefly introduced in next section and discussed in more details elsewhere [71].

## III. Theoretical Description

In this section, we introduce the reader to the different regimes of the spatiotemporal dynamics of the Omnipolarizer. We indicate with $\mathbf{S}=[S_1,S_2,S_3]$ and $\mathbf{J}=[J_1,J_2,J_3]$ the Stokes vectors associated with the forward and backward beams, respectively, whereas **s**=**S**/|**S**| and **j**=**J**/|**J**| denote the corresponding unitary Stokes vectors which define the SOPs over the unit radius Poincaré sphere. The spatiotemporal dynamics of **S** and **J** in the fiber is ruled by the following coupled nonlinear partial differential equations [66]:

$$c^{-1}\partial_t \mathbf{S} + \partial_z \mathbf{S} = \mathbf{S} \times \mathbf{DJ} - \alpha \mathbf{S}$$
$$c^{-1}\partial_t \mathbf{J} - \partial_z \mathbf{J} = \mathbf{J} \times \mathbf{DS} - \alpha \mathbf{J} \quad (1)$$

where D=$\gamma$ diag(-8/9,-8/9,8/9) is a diagonal matrix, $\gamma$ and $\alpha$ are the nonlinear Kerr coefficient and the linear propagation loss coefficient of the fiber, respectively, and $c$ is the speed of light in the fiber. With this notation, the component $s_3$ represents the forward signal ellipticity. In addition, in the following we indicate with R the 3x3 rotation matrix modeling the polarization rotation induced by the reflective-loop, and with $\rho$= |**J**(z=L,t)| / |**S**(z=L,t)| the output power ratio.

Along with Eq. (1), the boundary condition at the fiber end, which reads as **J**(z=L,t)=$\rho$R**S**(z=L,t) [66], allows to univocally determine **S**(z,t) and **J**(z,t) once that the input fields **S**(z=0,t) and **J**(z=0,t) are known. We point out that chromatic dispersion is neglected in our model, since the typical dispersion lengths of pulses propagating in the fiber is much longer than the fiber itself.

From Eqs. (1), we can see that the intensity-shape of |**S**| and |**J**| is preserved. Indeed |**S**| and |**J**| propagate unaltered except for linear propagation losses and a temporal shift, that is to say, |**S**(z,t)|= |**S**(0,t-z/c)|exp(-$\alpha$z) and |**J**(z,t)|= |**J**(0,t+z/c)|exp($\alpha$z-$\alpha$L). Intensity-shape preservation is an important feature of the Omnipolarizer, as it prevents input temporal polarization fluctuations to translate into large output intensity variations, which are referred as relative intensity noise (RIN) in literature. For this reason we define the Omnipolarizer as a *lossless* polarizer, differently from typical linear polarizers which suffer of RIN because of their nature of dissipative polarizers.

The dynamics of the Omnipolarizer is closely related to the stability of the stationary solutions of Eqs. (1), which are found by dropping the time derivatives. When the powers of the forward and backward beams are similar, which is the case for the bistable and alignment regimes, only the stable stationary SOPs, here indicated with **s**$_{stat}$ and **j**$_{stat}$, can play the role of polarization attractors for the output forward and backward signals. As a rule of thumb, it has been found that stable stationary states are characterized by a non-oscillatory behavior along the fiber length, while unstable states are oscillating [60]. This stability criterion is illustrated here in Fig. 2. If we numerically solve Eqs. (1) employing a stable-stationary state **s**$_{stat}$(z) as input longitudinal field, that is **s**(z,t=0) = **s**$_{stat}$(z), then this state is preserved in time due to the stability. Even if the input longitudinal field is slightly perturbed by some additional noise **n**(z), that is **s(z,t=0)** = **s**$_{stat}$(z) + **n**(z) (see Fig. 2a, red dotted line), however the stable stationary state plays the role of attractor which "cancels" the noise component, so that the field relaxes in time toward the original stable state, that is **s(z,t)** = **s**$_{stat}$(z) (see Fig. 2a, black solid line) for t>>1. To the opposite, if a stationary state characterized by several spatial periods along the fiber length, is now used as input longitudinal field (see Fig. 2b, red dotted line) then it has been found to be unstable. Indeed, this stationary state cannot be preserved in time: the system gradually leaves this state and converges towards a stable stationary state with a non-oscillatory behavior (see Fig. 2b, black solid line).

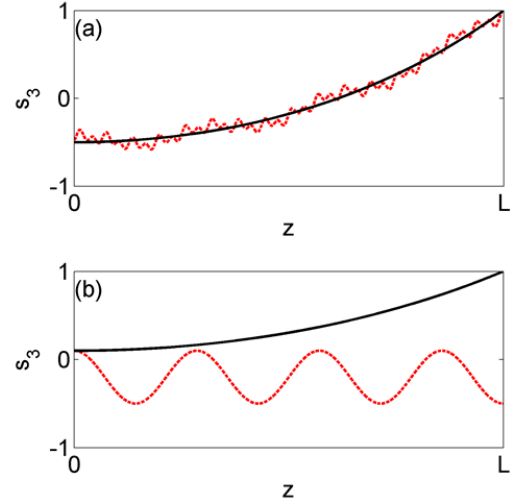

Fig. 2. Illustration of the stability criterion. Only the $s_3$ component is represented, $s_1$ and $s_2$ follow a similar behavior (a) A stable-stationary state, characterized by a non-oscillatory behavior along the fiber length, is employed as input longitudinal field with the addition of a weak noise component (red dotted line). Despite the initial presence of noise, the system converges in time toward the original stationary-state (black solid line). (b) An unstable-stationary state, characterized by an oscillatory behavior along the fiber length, is employed as input longitudinal field (red dotted line). The system leaves this unstable state and converges in time toward the stable state (black dotted line).

Figure 3 displays qualitatively the relation between the Omnipolarizer behavior and the stability of its stationary states (here depicted in circles). In panels (a-c) the bistable regime ($0.8 \leq \rho \leq 1$) is illustrated. The evolution along the fiber of the component $s_3$ of **s** is depicted by a solid line at 3 consecutive times, $t_A$<L/c, $t_B$>L/c and $t_C$>>L/c, where L/c defines the time-of-flight from one end to the other of the fiber. Let us assume, for the sake of simplicity, that the input signal is constant in time, i.e. **s**(z=0,t)≡**s**(z=0). At $t_A$<L/c (panel a) the backward replica has not been generated, yet, therefore **S** is not coupled to **J** and it propagates unchanged into the fiber. On the other hand, at $t_B$>L/c (panel b), **J** has been generated and it nonlinearly interacts with **S**, so that **s** gradually converges towards the stable state **s**$_{stat}$. The asymptotic convergence to the steady state forces **s**(z, t=$t_C$) and **s**$_{stat}$(z) to practically coincide ($s_3$(z,t=$t_C$) ≃ $s_{3,stat}$(z) for large times $t_C$>>L/c (see panel 3c).

We highlight that there is an unique stable stationary state **s**$_{stat}$(z) associated with a given input value **s**$_{stat}$(z=0), therefore **s**(z,t) converges towards the stable stationary state such that **s**$_{stat}$(z=0) = **s**(z=0). In addition, the stable stationary state basically depends on the reflection coefficient $\rho$ and on the total number N=L/L$_{nl}$ of nonlinear lengths. Note that N defines the degree of nonlinearity of the system: when N>>1 the Omnipolarizer exhibits a strong nonlinear dynamics such that the attraction processes previously outlined are truly efficient. In practice, $N \geq 4$ is usually sufficient to observe a strong attraction both in the bistable and in the alignment regime, and we thus refer to this condition as highly-nonlinear regime.

Interestingly enough, in the highly-nonlinear regime and when $0.8 \leq \rho \leq 1$, we observe that the output SOP $s(z=L,t)$ is asymptotically attracted towards only one of the poles of the Poincaré sphere: such pole depends on the sign of the input SOP ellipticity [66], namely, $s_3(z=L,t) \simeq \text{sign}[s_3(z=0)]$. Therefore, the value of $s_3$ at the output of the fiber approaches +1 or -1, simply depending on its input value $s_3(z=0)$. For this reason, a slight different input condition of $s_3(z=0)$ (leading to a change of sign of the input ellipticity) may lead to an output SOP which is switched to the orthogonal output polarization, which is the case displayed in Fig. 3a-c, where if $s_3(z=0)=+0.1$ then $s_3(z=L,t=t_C)\simeq+1$, otherwise if $s_3(z=0)=-0.1$ then $s_3(z=L,t=t_C)\simeq-1$.

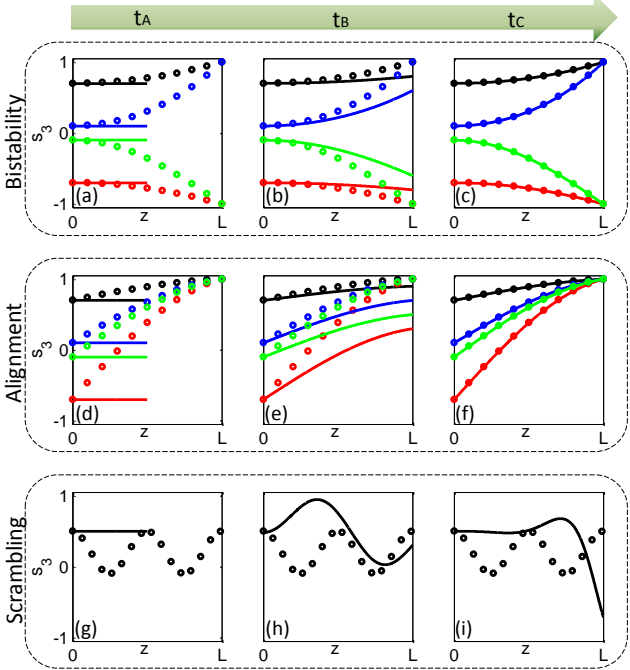

Fig. 3. Spatial evolution along the fiber length of the normalized Stokes component $s_3$ (solid lines) in the 3 regimes of the Omnipolarizer. Three consecutive instants are represented, which are $t_A<L/c$ (left panels a,d,g); $t_B>L/c$ (central panels b,e,h); $t_C>>L/c$ (right panels c,f,i). Corresponding stationary solutions $s_{3,stat}$ are represented in circles. Panels a-c: bistable regime. Four cases are represented, each one corresponding to different input polarizations $s_3(z=0)=+0.7$ (black lines and circles), $s_3(z=0)=+0.1$ (blue lines and circles), $s_3(z=0)=-0.1$ (green lines and circles), $s_3(z=0)=-0.7$ (red lines and circles). In all cases $s_3$ is asymptotically attracted in time towards the corresponding stable stationary solution $s_{3,stat}$. Panels d-f: same as in panels a-c but in the case of the attraction regime. In general, the attraction SOP at the fiber output is $s_{3,stat}(z=L) =+1$. In general, it is tunable by means of R and $\rho$. Panels g-i: scrambling regime. The stationary solution is unstable, consequently $s_3$ does not converge towards $s_{3,stat}$. The SOP fluctuates in time without reaching a fixed state and it thus turns out to be temporally scrambled at fiber output.

Panels (d-f) in Fig. 3 illustrate the alignment regime ($1.2 \leq \rho \leq 2$). As it occurs in the bistable regime, we observe again an asymptotic convergence of $s$ towards the stable stationary state $s_{stat}$ such that $s_{stat}(z=0) = s(z=0)$. However, due to a symmetry breaking induced by the rotation matrix R, one of the two poles of attraction becomes unstable, and only a single stable point of attraction survives. The Omnipolarizer then acts as a strong polarization funnel: whatever the input SOP is, we observe a unique attraction SOP at the fiber output which is tunable all over the Poincaré sphere by adjusting the rotation R and the coefficient $\rho$ [66].

Contrary to the bistable and the alignment regimes, when $\rho>>1$ the system is characterized by stationary states that are oscillating along the fiber length: this means that the steady states are unstable and cannot play the role of attractors anymore. Therefore, $s$ does not converge towards $s_{stat}$ in time, but it varies in every point of the fiber without reaching a fixed state. As a result, the signal SOP thus turns out to be temporally scrambled at the fiber output (panels g-i of Fig. 3).

## IV. EXPERIMENTAL SETUP

Figure 4 depicts the experimental setup implemented in order to characterize both operating regimes of the Omnipolarizer described above in the context of a 40-Gbit/s OOK Return-to-Zero (RZ) transmission. An initial 40-Gbit/s RZ signal is first generated thanks to 10-GHz mode-locked fiber laser delivering 2.5-ps pulses at 1562.4 nm. The pulses are then reshaped into 7.5-ps Gaussian pulses by means of a programmable optical bandpass filter. The resulting 10-GHz pulse train is intensity modulated by means of a LiNbO3 modulator driven by a $2^{31}-1$ pseudo-random bit sequence (PRBS) and finally multiplexed in the time domain so as to achieve a total bit rate of 40-Gbit/s. A commercial polarization scrambler is used to introduce strong polarization fluctuations at the input of the Omnipolarizer with a rate of 0.625 kHz. Before injection into the device, the 40-Gbit/s signal is amplified thanks to an Erbium doped fiber amplifier (EDFA) to an average power of 27 dBm.

The Omnipolarizer is made of a 6.2-km long Non-Zero Dispersion-Shifted Fiber (NZDSF) characterized by a chromatic dispersion D = -1.5 ps/nm/km at 1550 nm, a Kerr coefficient $\gamma=1.7$ $W^{-1}.km^{-1}$ and a PMD coefficient of 0.05 $ps/km^{1/2}$. An optical circulator is inserted at the input of the fiber in order to inject the incident signal as well as to extract the residual counter-propagating replica through its port #3. At the opposite end of the fiber, the feedback loop consists in a second circulator, a 90:10 coupler and a second EDFA whose gain enables us to directly adjust the reflection coefficient $\rho$. Note that a polarization controller is also inserted into the loop in order to select the orthogonal basis of SOP attractors over the Poincaré sphere.

After propagation, the repolarized 40-Gbit/s data signal was optically filtered by means of a 70-GHz Gaussian shape optical bandpass filter shifted by 170 GHz from the initial signal frequency, in order to reshape the output signal and improve its extinction ratio, just like the well-known Mamyshev regenerator [72-76]. At the receiver, the polarization state of the resulting 40-Gbit/s signal is characterized over the Poincaré sphere by means of a commercially available polarimeter. The residual polarization fluctuations as well as the intensity profile degradations are also characterized in the temporal domain by monitoring the corresponding eye-diagram beyond an inline polarizer.

Finally, the data are detected by means of a 70-GHz photodiode and electrically demultiplexed into the time domain to four 10-Gbit/s tributary channels to perform bit-error-rate (BER) measurements.

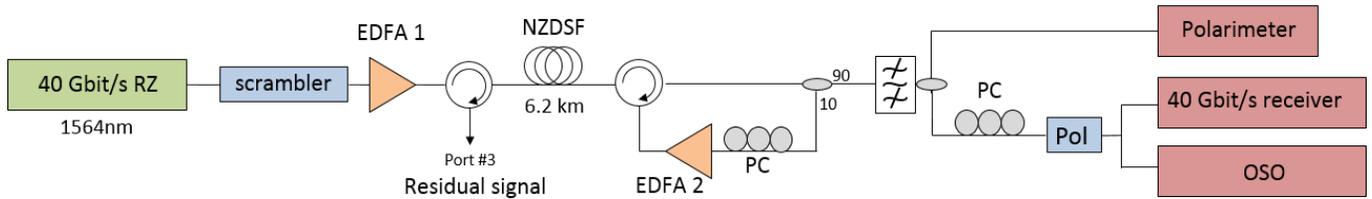

Fig. 4. Experimental setup of the Omnipolarizer: polarizer (Pol), Erbium doped fiber amplifier (EDFA), non-zero dispersion fiber (NZDSF), polarization controller (PC), Optical sampling oscilloscope (OSO).

## V. EXPERIMENTAL RESULTS

The efficiency of polarization attraction obtained by means of the Omnipolarizer was first characterized in the polarization domain. Figure 5a displays on the Poincaré sphere the SOP of the 40-Gbit/s signal recorded at the input of the device. Because of the initial polarization scrambling, the signal SOP distribution well covers the entire sphere. This arbitrarily polarized signal is then injected into the Omnipolarizer with an average power of 27 dBm. Figure 5b&c show the evolution of the SOP recorded at the output of the Omnipolarizer. In a first step (Fig. 5b), the back-reflected power is adjusted to an average power of 25 dBm by means of the EDFA 2, corresponding to a reflective coefficient $\rho=0.8$. In this case, and in agreement with the previous theoretical description, the Omnipolarizer enters into the bistable regime and the output SOP of the 40-Gbit/s signal converges to either one of two orthogonal SOPs, here adjusted on both poles of the sphere by means of the polarization controller of the reflected loop. In this regime, the Omnipolarizer clearly acts as a polarization beam splitter. Indeed, depending only on the sign of its initial ellipticity, all of the energy of the incident 40-Gbit/s signal is routed to either the north or south pole of the Poincaré sphere. It is noteworthy that these two output SOPs are independent of any environmental changes or stress induced on the fiber. Indeed, we checked that straining the fiber does not influence the position and width of the output SOP distribution.

Next, increasing further the backward power to 28 dBm ($\rho=1.2$) forces the Omnipolarizer to switch to the second regime of polarization attraction. In this regime, one of the previous two poles of attraction becomes unstable, and the device acts as a nonlinear polarizer. All input polarization fluctuations are canceled by the device and consequently, all output SOPs of the 40-Gbit/s signal remain trapped around a small area on the Poincaré sphere, as shown by Fig. 5c. Note that the selection of one or the other poles of attraction is only determined by the angle of polarization rotation in the reflective loop, and thus can be selected by means of the polarization controller. Such angle is at the origin of the symmetry breaking between the two spots of attraction [60].

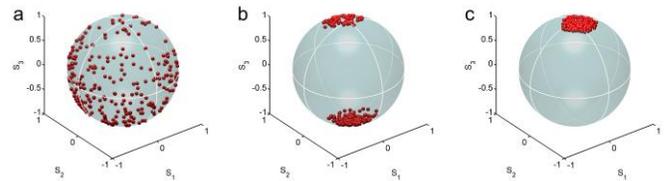

Fig. 5. (a) Poincaré sphere of the 40-Gbit/s signal at the input of the device (b) Poincaré sphere recorded at the output of the Omnipolarizer in the bistable regime ($\rho=0.8$) for an input power of 27 dBm (c) Poincaré sphere recorded at the output of the Omnipolarizer in the attraction regime ($\rho=1.2$)

The efficiency of this repolarization processing is even more impressive when monitored into the time domain. To this aim, we first display in Fig. 6a the eye-diagram of the 40-Gbit/s RZ signal recorded at the input of the Omnipolarizer and detected after an inline polarizer. Obviously, due to the initial polarization scrambling, all polarization fluctuations imposed by the scrambler are transferred into the time domain, thus inducing large intensity variations and a complete closure of the eye-diagram.

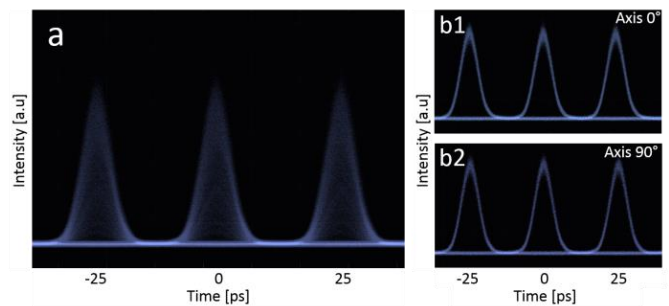

Fig. 6. (a) Eye-diagram of the 40-Gbit/s signal monitored behind an inline polarizer at the input of the device (b1-b2) Eye-diagram of the 40-Gbit/s signal at the output of the Omnipolarizer in the bistable regime and monitored along both axes of a PBS.

To the opposite, consider a signal that is injected into the Omnipolarizer in its bistable regime ($\rho=0.8$), corresponding to the output Poincaré sphere of Fig. 5b. In this case, as highlighted by figures 6b1 & b2, no intensity fluctuations can be observed on the output eye-diagrams of the 40-Gbit/s signal when detected behind a polarization beam splitter (PBS). Indeed, in this operating regime, the whole energy of the 40-Gbit/s signal can be routed in a binary manner to either one of the other axis of the output PBS depending only on the sign of its input ellipticity. The output eye-diagrams remain wide-open without any pulse splitting and with an extinction ratio above 20 dB between both axes, thus confirming the previous results observed on the Poincaré sphere in Fig. 5b.

The binary nature of the Omnipolarizer response, leading to switching of the output SOP among orthogonal axes, can be also clearly observed in Fig. 7. More precisely, we have recorded the 40-Gbit/s data stream at the output of the device and after a PBS by means of a 1-GHz low bandwidth oscilloscope, in such a way to only monitor the polarization fluctuations. These results confirm that, despite the initial polarization scrambling imposed on the input signal at a rate of 0.625 kHz, the output SOP has a binary nature, since it is routed to either one or the other axis of the PBS with a high extinction ratio. This process could be useful for instance to develop an all-optical random bit generator [77] from a chaotic polarization device, as well as a polarization-based optical memory or a polarization switch [68].

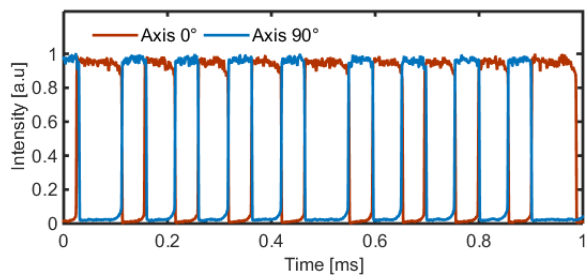

Fig. 7. Temporal evolution of the polarization state of the 40-Gbit/s signal recorded at the output of the Omnipolarizer in the bistable regime and monitored along both axes of a PBS.

By increasing further the average power of the backward replica until 28 dBm by means of the EDFA 2, the Omnipolarizer switches to its attraction regime, corresponding to the Poincaré sphere reported in Fig. 5c. As in the previous study, the eye-diagram of the 40-Gbit/s signal is then recorded at the output of the device after an inline polarizer in order to transfer the residual polarization fluctuations into intensity noise. The resulting eye-diagram is depicted in Fig. 8a. Outstandingly, despite the input polarization scrambling, illustrated by the initial closed eye-diagram of Fig. 6a, here we can clearly observe through the output polarizer a well-open eye-diagram. Indeed, no intensity fluctuations can be observed beyond the polarizer, which confirms the efficiency of the self-induced repolarization process undergone by the signal within the Omnipolarizer.

More importantly, as displayed in Fig. 8b, the corresponding bit-error-rate measurements, performed after a polarizer as a function of the incoming power on the receiver, show that the Omnipolarizer is able to maintain the maximum of performance of data-processing through any polarization-dependent transmission component. Indeed, whereas the input BER is limited to a threshold of $10^{-2}$ due to the initial scrambling process which closes the eye-diagram through the polarizer, the output signal is forced to align its SOP on the polarizer axis, thus enabling a complete recovery of the transmitted data with an error-free measurement for all the four 10-Gbit/s tributary channels demultiplexed from the 40-Gbits data stream.

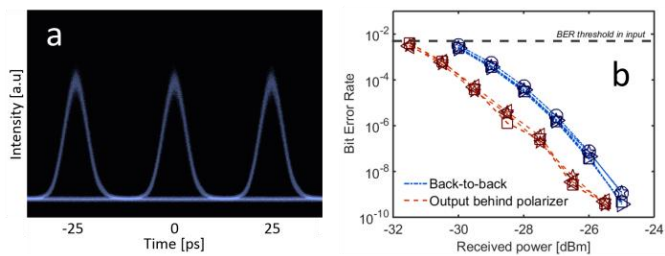

Fig. 8. (a) Eye-diagram of the 40-Gbit/s signal at the output of the Omnipolarizer in the attraction regime and monitored behind a polarizer (b) Corresponding bit-error-rate measurements as a function of received power for the four 10-Gbit/s tributary channels demultiplexed from the 40-Gbit/s signal.

It is also worth noting that the output BER is characterized by a 1 dB sensitivity improvement with respect to back-to-back measurements. This phenomenon is attributed to the pulse reshaping properties of the device. Indeed, to be efficient, the polarization attraction regime of the Omnipolarizer imposes a high level of average power close to 500 mW. Consequently, the optical pulses of the RZ 40-Gbit/s signal propagate in the NZDSF with a peak power around 3 W, and thus undergo a large self-phase modulation (SPM) effect. The SPM leads to a wide spectral broadening, as illustrated in Fig. 9a (orange curve) after the NZDSF, when compared with the input spectrum (blue curve). This SPM effect combined with the normal chromatic dispersion of the fiber induces a nonlinear reshaping of the pulses towards a parabolic shape first and then to a square shape [78], as shown in Fig. 9b with the eye-diagram recorded at the output of the NZDSF. In order to retrieve the initial shape of the pulses and improve the extinction ratio, we thus carried out a 170-GHz Gaussian shape offset filtering (see output spectrum in Fig. 9a, red solid-line), which led us to obtain the eye-diagram of Fig. 8a, by using the principle of a Mamyshev regenerator [72-76]. Therefore, it is important to note that the strong nonlinear regime of propagation undergone in the Omnipolarizer can also be a limitation for practical implementations to high-bit-rate signals. Indeed, a careful design, in particular a proper choice of the fiber has to be done, especially for OOK signals, in order to avoid complex soliton dynamics in anomalous dispersive fibers [69], or wave breaking phenomena in the normal dispersion regime [69, 79].

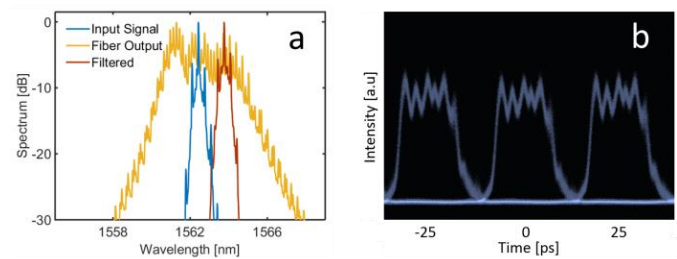

Fig. 9. (a) Optical spectra of the 40-Gbit/s signal measured at the input of the device (in blue), at the output of the NZDSF fiber (in orange) and after the offset filtering (red solid-line) (b) Eye-diagram of the 40-Gbit/s signal directly recorded at the output of the NZDSF fiber in the attraction regime (P=27 dBm) and monitored without filtering operation.

## VI. POLARIZATION-BASED TUNNELING EFFECT

An important physical aspect of the Omnipolarizer lies in the fact that in the nonlinear regime it can reach a stable stationary state where both the forward and backward waves are coupled in each and every point along the fiber. By analogy with the case of two parallel pendulum chains of magnets which are coupled at every point of propagation, here any fluctuation of a wave along the fiber length is immediately inscribed onto the other wave.

This remarkable fact can have important consequences. A fascinating example is that, in the strong nonlinear regime, the previously discussed self-induced attraction process allows to all-optically speed-up the propagation of a polarization burst within the system in such a way to instantaneously write, as a sort of nonlinear mirror, a polarization information onto the backward signal, long before the linear round-trip time of light in the fiber.

To this aim, the Omnipolarizer was first subjected to its stationary regime thanks to the injection of a CW signal with a fixed average power P and SOP. Next, short polarization spikes are imprinted onto the input CWs by means of our polarization scrambler/controller in such a way to impose fast and brutal SOP variations onto the Poincaré sphere. More precisely, these incident fast polarization pulses consist of $S_3$ spikes whose time duration has been adjusted to 10 µs.

First of all, the polarization burst is injected into the Omnipolarizer in such a way to propagate in a quasi linear regime. The input average power if then fixed to 16 dBm and $\rho$ close to 0.8. The temporal profile of this polarization burst is then detected after a polarizer at each port of the device, i.e. input, output as well as in port #3 of the input circulator (residual signal in Fig. 4) to measure the backward signal. Figure 10a displays the experimental monitoring of this polarization burst propagating within the Omnipolarizer in the linear regime. At the origin of times, the polarization burst enters into the system (blue solid-line). Then, this polarization impulse is detected at the output of the device at the time corresponding to the time-of-flight along the NZDSF span, here 31 µs (red solid-line). Finally, after a reflection at the fiber output and back-propagation, the polarization spike is basically detected on the oscilloscope after a delay corresponding to a linear round-trip in the fiber, i.e. 62 µs (orange solid-line). We thus denote with "classic-spike" this usual backward replica.

This usual situation turns to be completely different when operating in the nonlinear regime. The input power is therefore increased up to 27 dBm whilst $\rho$ is kept constant close to 0.8. For that power, the Omnipolarizer is in a strong nonlinear regime. The experimental monitoring of the polarization burst transmission is now displayed in Fig. 10b and reveals a nontrivial behavior. Whereas the temporal profiles recorded at the input and output of the fiber remain similar to the previous linear regime, the reflected replica now clearly exhibits the signature of the polarization spike long before its classical round-trip along the fiber span. We talk in this case of "early-spike", as it is marked in Fig. 10b. In fact, in analogy with the quantum tunneling mechanism, here the data carried by the polarization state of the signal has temporally "jumped" from the input to the output, without a delay imposed by its finite speed of propagation. As a matter of fact, the early-spike seems to exit from the "residual signal" port#3 of the input circulator at the same time that the forward input spike is injected into the fiber, similarly to a fast-light phenomenon [85].

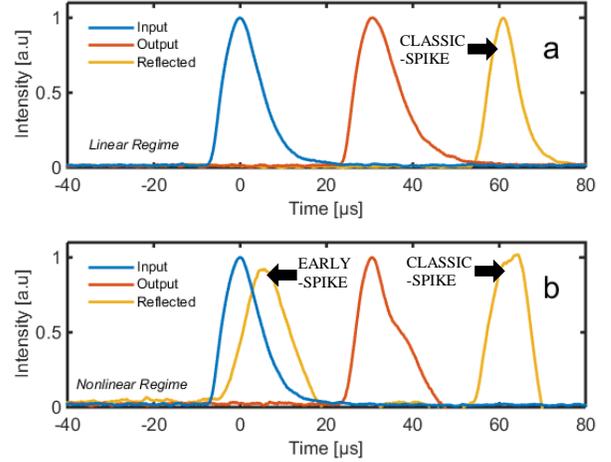

Fig. 10 (a) Intensity profile of the polarization burst propagating into the Omnipolarizer (a) Linear regime, P=16 dBm (b) Nonlinear regime, P=27 dBm.

As previously anticipated, this tunneling effect is ascribable to the local coupling between the forward and backward waves in the fiber: for this reason the stronger the effective device nonlinearity, the larger the intensity of the early-spike. In a linear regime the early-spike does not appear, instead, because forward and backward waves are not coupled to each other, which is the case illustrated in Fig. 10a.

Numerical simulations of this phenomenon have been carried out based on Eqs. (1) and confirm our previous experimental observations. Numerical results are then displayed in Fig. 11 and illustrate the temporal profile of a single polarization burst recorded at each port of the device, i.e. the input forward burst (blue line), the output forward burst (red line) as well as the backward burst replica (green line). The parameters used in our simulations correspond to the experimental configuration reported in Fig. 10. In the case of panel 11a, the power of the CW input signal, over which the input polarization burst is imprinted, is 16 dBm. With such a low power, nonlinear effects do not play any substantial role and the system is thus in a quasi-linear regime. Therefore, the input burst is detected at the output of the fiber at a time corresponding to the time-of-flight, while its usual back-reflected replica, namely the "classic-spike" in green line is well detected at port #3 of the Omnipolarizer after a delay of twice the time-of-flight, just as in experimental results of Fig. 10a.

This situation turns to be completely different when we increase the input power. Panel 11b displays the temporal profiles of the polarization burst when the initial power P is increased up to 24 dBm. The reflected replica now clearly exhibits the signature of the polarization burst characterized by the appearance of an early-spike long before its classical round trip along the fiber span. Finally, to prove that the "early-spike effect" is linked to the system nonlinearity, we further increase the power P up to 27 dBm in panel 11c. In this strong nonlinear regime, it is noteworthy that the early-spike rises even higher than the classic-spike. Note also, in panels 11(b&c), that the shape of both the classic-spike in the backward replica as well as the output forward polarization burst are both degraded by system nonlinearity.

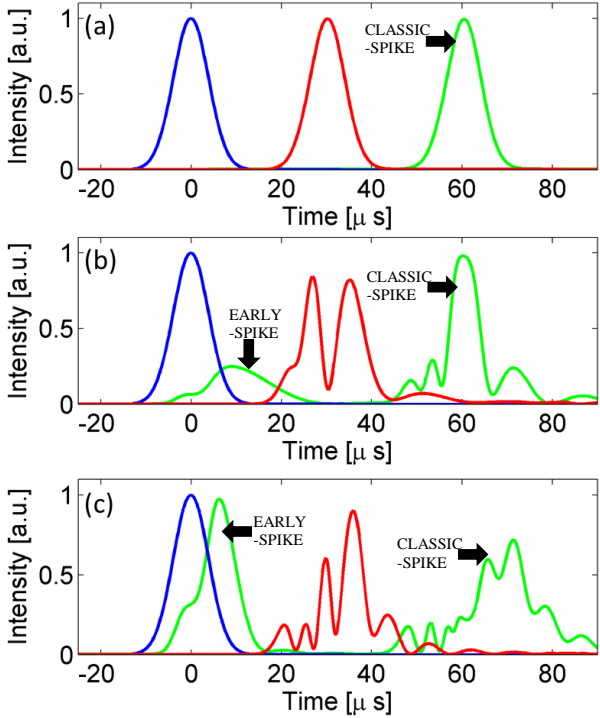

Fig. 11 Numerical simulations illustrating the intensity profile of the polarization burst propagating into the Omnipolarizer: input forward burst (blue line), output forward burst (red-line) and backward burst replica (green-line). (a) Linear regime, P=16 dBm (b) Nonlinear regime, P=24 dBm (c) Strong nonlinear regime, P=27 dBm. Other parameters correspond to the experimental configuration of Fig. 10.

It is also important to notice that the nonlinear response time of the Omnipolarizer, which is of the order of $L_{nl}/c$ [66], here close to 5 µs, fixes a lower bound to the time width $T_{spike}$ of the forward input polarization burst which can be detected. Indeed, if $T_{spike} \ll L_{nl}/c$ then the forward polarization burst is "transparent" to the nonlinear dynamics of the Omnipolarizer and it does not give rise to any backward early-spike: in such an instance only the classic-spike appears, as if the Omnipolarizer worked in a linear regime.

It is worth noting that a train of forward input polarization spikes carrying optical data could be injected: in this case a corresponding train of early-spikes would be efficiently copied and detected on the backward wave beyond the port #3 of the input circulator of the Omnipolarizer. The illustration of that process is depicted in Fig. 12 by means of numerical simulations. Note that in this case, the last early-spike should appear before the first classic-spike, otherwise they would overlap and the early-spike would be lost. Being L/c the round-trip time at which the first classic-spike is formed, we have thus at disposition a temporal transmission window of duration L/c during which a train of forward input spikes is quasi-instantaneously copied in the backward replica, as it is displayed in Fig. 12.

In the simulation under analysis we employ the 6-km long NZDSF, therefore the round-trip time is 60 µs. The input power P is increased up to 36 dBm, so that the nonlinear response $L_{nl}/c \approx 0.5$ µs. This permits us to employ bursts whose width $T_{spike} = 1$ µs, so that several input bursts can be injected in the fiber giving rise to their corresponding early-spikes.

We conclude this section by highlighting that the polarization of the backward bursts depends on the stationary-state which is reached by the system before the burst injection. Therefore, we may have input forward bursts, which are linearly polarized along the y-axis and corresponding early-spike replicas which are polarized along another direction, let-us say the x-axis. This paves the way to important potential applications in the framework of data-processing. Indeed, from the example discussed above it is clear that a train of y-polarized input forward bursts may be quasi-instantaneously transcribed into a train of x-polarized backward early-spikes, which would give rise to an ultrafast polarization converter.

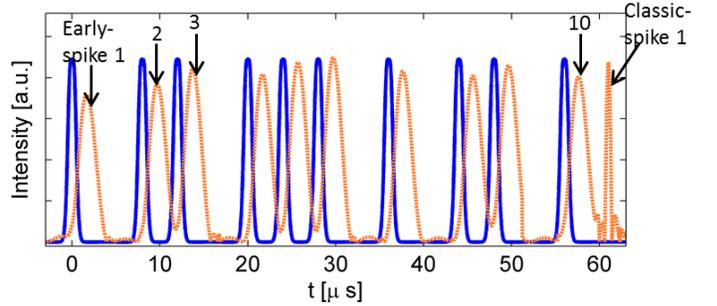

Fig. 12 Numerical simulation displaying the copying process and efficient detection of 10 early-spikes carrying a data packet. The round-trip time of the fiber is here 60 µs. A train of 10 forward input polarization spikes (blue solid line) is injected into the fiber and is quasi-instantaneously copied (orange dotted line) on the backward replica and then detected beyond the port #3 of the input circulator. The first three early-spikes are indicated along with the last one (10), which is formed just before the first classic-spike appears at t=60 µs (round-trip time): all the 10 early-spikes are thus correctly detected.

## VII. SELF-INDUCED MODAL ATTRACTION

The concept of self-induced polarization attraction within the Omnipolarizer discussed in Secs. II-V may be extended to the spatial modes of a multimode fiber. This paves the way to the more general concept of *self-induced modal attraction*, where the modes may be either polarization or spatial modes of a fiber. A new scenario emerges, where light could self-organize its modal state, namely, the power distribution among the modes as well as their relative phase.

A plethora of paramount applications could benefit from such a phenomenon of modal self-organization. Self-induced modal attraction could lead to the development of an all-optical signal processing technology in the framework of spatial-division-multiplexing schemes, which are rapidly emerging as the most promising solution in order to face the upcoming capacity crunch of current single-mode fiber systems [86]. Furthermore, it could provide an efficient and all-optical way to counteract parasitic and uncontrolled modal coupling effects in short multimode fibers, which are nowadays widely employed in several contexts, e.g., for carrying a large amount of data
8

from large-area telescopes to spectrometers. Finally, self-induced modal attraction could be exploited in multicore fiber lasers for the phase-synchronization of the cores, which allows for focusing most of the energy in the fundamental supermode of the fiber: such an idea has already been explored to synchronize some tens of cores [87], but a clear theoretical understanding is still missing and could be the key to synchronize an unprecedented number of cores.

The aim of this section is not to provide a comprehensive picture of modal attraction, which is as much intriguing as challenging and will deserve deep investigations in the near future. Here we want to discuss a simple but clear example of modal attraction which permits to disclose its potential application.

We warn the reader that the CNLSEs ruling the dynamics of spatial modes in a telecom fiber are complex to derive even in the simple bimodal case, which is due to the presence of a randomly varying birefringence along with polarization and spatial mode dispersion effects [52]. Therefore, for the sake of simplicity, here we consider a bimodal isotropic fiber where the two spatial modes co-propagate and interact with their backward replica generated by a perfectly reflecting mirror. A schematic setup of the device under analysis is illustrated in Fig. 13.

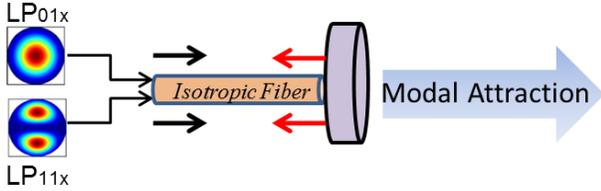

Fig. 13: Schematic illustration of the setup undergoing self-induced modal attraction. A combination of $LP_{01x}$ and $LP_{11x}$ modes is injected into a bimodal isotropic fiber (black arrows) and nonlinearly interacts with its backward replica (red arrows) generated at the fiber output by means of a reflecting device.

We indicate with $F_n$ and $B_n$ the electric field envelopes of the forward and backward mode of order-$n$, respectively. Assuming that all the modes are polarized along the same direction, e.g. the x-axis, and following a treatment similar to that discussed in [65], we find that the spatiotemporal dynamics of mode coupling is described by the following CNLSEs:

$$c_n^{-1}\partial_t F_n + \partial_z F_n = 2C_{01}F_m B_m B_n^* + \left(C_{nn}|F_n|^2 + 2C_{01}|F_m|^2 + 2C_{nn}|B_n|^2 + 2C_{01}|B_m|^2\right)F_n \quad (2)$$

where n={1,2}, m={1,2}, n≠m, whereas $c_n$ is the group velocity related to mode-$n$. A similar equation is found for $F_n$ by exchanging F↔B and $\partial_z$ ↔ - $\partial_z$. Here, differently from Ref. [65], we make the realistic assumption that the nonlinear coefficients $C_{00}$, $C_{11}$ and $C_{01}$, defined as in Ref. [88], may not be equal and we employ the boundary condition $B_n = -F_n$ at the fiber end, which accounts for the presence of the perfect mirror. Note that $C_{00}$ plays the role of the Kerr nonlinear coefficient $\gamma$ in the previously discussed single-mode fibers. Furthermore, we point out that the term $2C_{01}F_m B_m B_n^*$ in Eqs (2) is responsible for the energy exchange between the two fiber modes, and that the strength of the exchange is thus proportional to $C_{01}$.

In analogy with polarization phenomena, we define the modal Stokes parameters $\mathbf{S^{(m)}}$ =[$F_1 F_2^*$+ $F_2 F_1^*$, $iF_1 F_2^*$ - $iF_2 F_1^*$, $|F_1|^2$-$|F_2|^2$] and $\mathbf{J^{(m)}}$ =[$B_1 B_2^*$+ $B_2 B_1^*$, $iB_1 B_2^*$ - $iB_2 B_1^*$, $|B_1|^2$-$|B_2|^2$], as well as the nonlinear length $L_{nl}=1/(C_{00}P)$ and the corresponding number of nonlinear lengths N=L/$L_{nl}$, being P the total forward power that in this case reads as P=|$\mathbf{S^{(m)}}$|=$|F_1|^2$+$|F_2|^2$. Note that we employ the superscript (m) to differentiate modal Stokes vectors from polarization ones. A modal Poincaré sphere, completely analogue to the polarization Poincaré sphere, can conveniently display the position of the normalized unitary modal Stokes vector $\mathbf{s^{(m)}}$ = $\mathbf{S^{(m)}}$ / |$\mathbf{S^{(m)}}$|.

We solve Eqs. (2) for N=4 and for different CW input conditions {$F_1$(z=0),$F_2$(z=0)} corresponding to an uniform coverage of the modal Poincaré sphere by $\mathbf{s^{(m)}}$(z=0), as depicted in Fig. 14a. We finally plot the corresponding output $\mathbf{s^{(m)}}$(z=L,$t_C$) (Figs. 14b, 15b) at a large time $t_C$ such that the system has relaxed towards an asymptotic stable stationary state of Eq.(2).

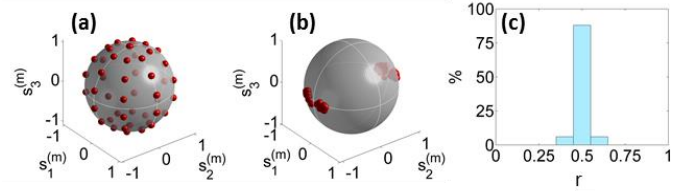

Fig.14: (a) and (b) input and output distribution of the modal Stokes vector $\mathbf{s^{(m)}}$ when L=4 and $C_{00}$=$C_{11}$=1.75$C_{01}$. (c) Histogram of $r$ at the fiber output: note that the central bin, corresponding to the range 0.45<$r$<0.55, represents more than the 80% of the total area of the histogram.

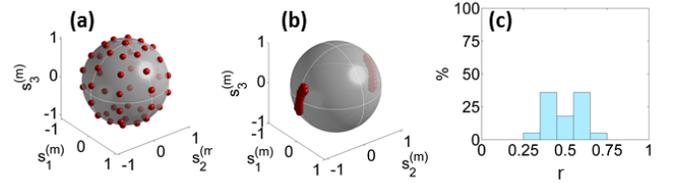

Fig.15: (a) same as in Fig. 14 but when $C_{00}$=$C_{11}$=3.5$C_{01}$. Note that $r$ is not strongly distributed around 0.5 as in Fig. 14, which is indicative of a weaker equipartition effect.

In Fig. 14b we observe that, whenever $C_{00}$=$C_{11}$=1.75$C_{01}$, a strong attraction is observed towards two distinct points of the sphere. Most important, all the vectors $\mathbf{s^{(m)}}$(z=L,$t_C$) lie close to the circle which corresponds to $s_3^{(m)}$=0, that is $|F_1|^2$-$|F_2|^2$=0. This means that whatever the input values $F_1$(z=0) and $F_2$(z=0) are, at the output of the fiber an equipartition of the energy among the two modes is observed, i.e. $|F_1|^2$≈$|F_2|^2$, in a similar way than in the *bistable* regime of the Omnipolarizer. The histogram of the ratio $r$ = $|F_1|^2$/( $|F_1|^2$+$|F_2|^2$ ) in Fig. 14c confirms the modal equipartition: at the fiber input $r$ is uniformly distributed between 0 and 1, whereas at the fiber output more than the 80% of the occurrences of $r$ lies in the range 0.45<$r$<0.55.

If we reduce the intermodal nonlinear coefficient $C_{01}$ so that $C_{00}$=$C_{11}$=3.5$C_{01}$, then the output modal equipartition is less efficient, as it is evident from Figs. 15b&c. This is not surprising, as the strength of the energy exchange between the two fiber modes is proportional to $C_{01}$: in the limit case $C_{01}$=0 no modal energy exchange occurs. As a result, the input and output distributions of $r$ are the same, namely, the effect of output energy equipartition is annihilated.

We point out that coefficients $C_{00}$, $C_{11}$ and $C_{01}$ can be adjusted over a wide range of values by means of an appropriate fiber design (see e.g. [88]). Therefore the two cases here analyzed, that is $C_{00}=C_{11}=1.75C_{01}$ and $C_{00}=C_{11}=3.5C_{01}$, are obtained for realistic fiber designs. It turns out that, in a strong nonlinear regime a bimodal fiber could be employed to achieve a truly efficient all-optical equipartition of energy among the two modes, whatever their input energy distribution is. Moreover, as in the Omnipolarizer, a reflective coefficient larger than 1, by means of a multimode amplified loop, should allow to completely trap the output energy in a single mode. From a general point of view, many other applications may be envisaged by exploiting the propagation of multiple modes. Indeed, the new field of all-optical modal attraction and self-organization is just at its beginning, and we expect it will provide exciting results in the near future.

## VIII. CONCLUSION

In this work, we presented a theoretical, numerical and experimental description of the self-induced repolarization process of light in optical fibers within a device called the Omnipolarizer. The principle of operation of this device is based on a counter-propagating cross-polarization interaction between a forward propagating signal and its backward replica generated at the fiber output by means of a reflective loop. We have successfully exploited the device in order to self-repolarize a 40-Gbit/s RZ OOK signal. Depending on the power ratio between the two counter-propagating waves, we have first been able to track the output polarization of the 40-Gbit/s signal in such a way to align its SOP on the axis of a polarizer, thus allowing for a polarization-independent error-free reception. Moreover, with a reflection coefficient just below unity, we observed the repolarization of light along two orthogonal output SOPs. The SOP at the output of the Omnipolarizer is therefore simply fixed by the sign of the ellipticity of the input signal. Indeed, experimental recordings of the 40-Gbit/s eye-diagrams confirm the binary nature of the output polarization state along two orthogonal channels with an extinction ratio higher than 20 dB. A fascinating physical aspect of the Omnipolarizer has also been exploited in order to demonstrate a polarization-based temporal tunneling effect. More precisely, the localized coupling induced by the cross-polarization interaction between the two counter-propagating waves allowed us to virtually instantaneously convert a polarization information onto the reflected signal, long before the expected linear time-of-flight induced by a reflective round-trip into the fiber. Finally, in the last section of this work, we presented the generalization of the idea of self-induced attraction to the case of modal attraction in multimode fibers. We discussed a simple but effective example of all-optical energy equipartition in a bimodal fiber, which is based on the nonlinear interaction between the two forward spatial modes of the fiber with their replicas generated by a reflection at the fiber output end. This simple example let us envisage a plethora of new applications based on self-organization phenomena in multimode fibers, which may find useful applications in the next generation of optical communication systems based on mode division multiplexing.

ACKNOWLEDGEMENT

This research was funded by the European Research Council under Grant Agreement 306633, ERC PETAL. https://www.facebook.com/petal.inside. We also thank the financial support of the Conseil Régional de Bourgogne in the frame work of the Photcom project, FEDER and the Labex ACTION (ANR-11-LABX-0001-01). The work of S. W. was supported by the Italian Ministry of University and Research (MIUR, Project Nb. 2012BFNWZ2). We thank Drs. D. Sugny, H. R. Jauslin, A. Picozzi, G. Millot and especially S. Pitois for fruitful discussions.